\documentclass[showpacs,prl,preprintnumbers,amsmath,amssymb]{revtex4}

\usepackage{graphicx}
\usepackage{dcolumn}
\usepackage{bm}

\begin{document}

\title{Inhomogeneous distribution particles in self-gravitational system.}

\author{B. I. Lev}

\affiliation{Bogolyubov Institute for Theoretical Physics, NAS of
Ukraine, Metrolohichna 14-b, Kyiv 03680, Ukraine}

\date{\today}

\begin{abstract}

The microcanonical partition function for self-gravitational
system in three dimensional case has been found. Used approach
from the field theory of statistical description of the system
was tailored to gravitational interacting particles with regard
for an arbitrary spatially inhomogeneous particle distribution.
The entropy of self-gravitational system has been found from
extreme condition for the effective functional. For inhomogeneous
distribution particle (formation few cluster of finite size) the
entropy are bigger as the entropy homogeneous distribution of
particles. The increasing of entropy of self-gravitational system
after formation cluster motive tendency to disperse.

\end{abstract}

\maketitle

The study of self-gravitational system has as fundamental as
practical physical interest. The self-gravitational system are
interesting for testing ideas about the statistical mechanic
description of systems governed by long range interaction.  A
self-gravitational system has also more general problem, studied
for a long time \cite{Pad}. The standard methods of statistical
mechanic cannot be carrier to study gravitational system. Due to
this fundamental difference, the notion of equilibrium is not
always well defined and those system exhibit a nontrivial
behaviour with gravitational collapse. For system with
gravitational interaction the thermodynamically ensemble are
inequivalent, negative specific heat \cite{Tir} in microcanonical
ensemble which not exist in canonical description \cite{Cha}.In
the microcanonical ensemble self collapse correspond a
"gravithermal catastrophe" and in canonical ensemble to an "
isothermal collapse" \cite{Cha}. A self gravitational system can
increase entropy without bound by developing a dense and hot core
surrounded by dilute halo. Since equilibrium states are only
local entropy maximum. However, if introduce a repulsive
potential at short distance, complete core collapse is prevented
and can to proved that a global entropy maximum now exist for all
accessible values of energy. The effective repulsion can be
introduce in many different way but the physical results are
rather insensitive to the precise form of regularization.
Alternative can consider a classical hard-sphere gas by
introducing excluded volume around each particle \cite{Aro}. For
the gas with pure gravitational interaction between its particles
the virial coefficients for potential of particles interaction
$1/r^{n}$ if $n<3$ and thus the partition function diverges, the
energy of gravitating particle will not be an extensive
parameter. The entropy such system tends to infinity too when
volume $V \to \infty $. In it was shown that a self-gravitating
gas collapses. A nature of the collapse and its conditions are
explained by using simple and clear consideration \cite{Lyn}. The
phase transition in such systems creates the problem of
description in the mean-field thermodynamics approach \cite{Cha}.
Two type approaches (statistical and thermodynamic) have been
develop to determination the equilibrium states of
self-gravitational system \cite{Cha}, \cite{Cav}. About all this
problem and possible solution very good described in review
\cite{Chav}. The collapse in such system are begin as spatially
homogeneous distribution of particles in the all system at once.
Formation of the spatially inhomogeneous distribution of
interaction particles is a typical problem in condensed matter
physics and requires non-conventional methods of statistical
description of the system was tailored to gravitational
interacting particles with regard for an arbitrary spatially
inhomogeneous particle distribution. This method must employs the
procedure to find dominant contributions to the partition
function and to avoid entropy divergences for infinite system
volume. Only a few model systems with interaction are known for
which the partition function can be exactly evaluated, at least
within thermodynamic limits \cite{Bax} but not for inhomogeneous
distribution of particle. As for existence of equilibrium states,
a few result have been obtained in the framework of "exact"
equilibrium statistic mechanics with one considered a finite
number of particles \cite{Aly}. Is no known exact solution in
even in three dimensional self- gravitational system. Formation
of the spatially inhomogeneous distribution of interaction
particles requires a nonconventional method, such as use in
\cite{Bel}, \cite{Lev}, \cite{Kle}, which are based on
Hubbard-Stratonovich representation of statistical sum
\cite{Str}. This method is now extended and applied to a
gravitational interacting system to find solution for particles
distribution without using spatial box restrictions. It is
important that this solution has no divergences in thermodynamic
limits. For this goal can use saddle point approximation which
take into account the conservation number of particles in
limiting space, which provided to nonlinear equation. Described
condition give the possibility determine microcanonical partition
function for self- gravitational system. This partition function
in the case homogeneous distribution of particle, and in case
inhomogeneous distribution with formation few equal cluster with
same size has not any peculiarity and fully determined
thermodynamic parameter of system. Has been shown, that the
entropy for inhomogeneous distribution of particle in
self-gravitational system are bigger as entropy homogeneous
distribution of particle. This thermodynamic force, which appear
by formation spatial inhomogeneous distribution of particle can be
motivation of motion of cluster.

In particular can consider the microcanonical evolution of
self-gravitational system. In the microcanonical ensemble, the
object of fundamental interest is the density of states, which
can written in the standard form \cite{Hua}
\begin{equation}
\Omega_{E,N}=\int d\mathbf{q}
d\mathbf{p}\delta(H(\mathbf{p},\mathbf{q})-E)\delta(N(\mathbf{p},\mathbf{q})-N)
\end{equation}
where $\Omega_{E,N}$ denote microcanonical partition function for
fixed energy and number of particles in system. Introduce the
Lagrange multiplier $\beta=\frac{1}{kT}$ inverse temperature  and
$\eta=\beta\mu$ where $\mu$-is chemical potential and make Laplase
transformation can present the microcanonical distribution
function in the form :
\begin{equation}
\Omega_{E,N}=\oint d\beta \oint d\eta \exp\left\{\beta E-\eta N
\right\}\int d\mathbf{q} d\mathbf{p}\exp\left\{-\beta
H(\mathbf{p},\mathbf{q})+\eta N(\mathbf{p},\mathbf{q}) \right\}
\end{equation}
The entropy and temperature for microcanonical ensemble are
defined by $S=\ln \Omega_{E,N}$ and inverse temperature can
determined from relation $\beta=\frac{d S}{dE}$. The pressure of
system defined through formula $P=\frac{1}{\beta}\frac{\partial
S}{\partial V}_{E}$ and present the equation of state for system.

A system of interacting particles can be treated in the classical
manner as Ising model with Hamiltonian \cite{Mag} present in term
occupation number in the form:

\begin{equation}
H(n)=\sum_{s}\varepsilon _{s}n_{s}-\frac{1}{2}\sum_{s,s^{\prime
}}W_{ss^{\prime }}n_{s}n_{s^{\prime }}
\end{equation}

where $\varepsilon _{s}$ is the additive part of the energy in
the state $s$ which is equal in most cases to the kinetic energy
\cite{Mag}, $W_{ss^{\prime }}$ are interaction energy for the
particles in the states $s$ and $s^{\prime }$. In this model the
macroscopic states of the system are described by a set of
occupation numbers $n_{s}$. Index $s$ labels an individual
particle state; and can correspond as well as a fixed site on the
Ising lattice \cite{Isi}, which explicit form is irrelevant in
the continuum approximation. The number of particle is fixed
which can determine from relation $N(n)=\sum_{s}n_{s}$. It is
clear that calculating of the partition function as a rather
complicated problem even in the case of the Ising model. The
microcanonical partition function of a system of interacting
particles can given in the form \cite{Lev}:
\begin{equation}
\Omega_{E,N}=\oint d\beta \oint d\eta \exp\left\{\beta E-\eta N
\right\} \sum_{\left\{ n\right\} }\exp \left( -\beta H(n)\right)
\end{equation}
or explicit definition form
\begin{equation}
\Omega_{E,N}=\oint d\beta \oint d\eta \exp\left\{\beta E-\eta N
\right\}\sum_{\left\{ n\right\} }\exp \left\{ -\beta \left[
\sum_{s}\varepsilon _{s}n_{s}-\frac{1}{ 2}\sum_{s,s^{\prime
}}W_{ss^{\prime }}n_{s}n_{s^{\prime }}\right] \right\} \label{3}
\end{equation}
where $\sum\limits_{\left\{ n\right\} }$ implies summation over
all probable distributions $\left\{ n_{s}\right\} $. In order to
perform a formal summation, additional field variables can be
introduced making use of the theory of Gaussian integrals
\cite{Str}, \cite{Mag}:
\begin{equation}
\exp \left\{ \frac{1}{2\theta }\nu ^{2}\sum_{s,s^{\prime }}\omega
_{ss^{\prime }}n_{s}n_{s^{\prime }}\right\} =
\int\limits_{-\infty }^{\infty }D\varphi \exp \left\{ \nu
\sum_{s}n_{s}\varphi _{s}-\frac{1 }{2\beta}\sum_{s,s^{\prime
}}\omega _{ss^{\prime }}^{-1}\varphi _{s}\varphi _{s^{\prime
}}\right\}
\end{equation}
where $D\varphi=\frac{\prod\limits_{s}d\varphi _{s} }{\sqrt{\det
2\pi \beta \omega _{ss^{\prime }}}}$ and $\omega _{ss^{\prime
}}^{-1}$ is the inverse of the interaction matrix. The latter
satisfies the condition $\omega _{ss^{\prime \prime }}^{-1}\omega
_{s^{\prime \prime }s^{\prime }}=\delta _{ss^{\prime }}$ with
$\nu ^{2}=\pm 1$ depending on the character of interaction
energy. If introduce instead of chemical potential other
important variable - chemical activity $\xi\equiv e^{\beta
\mu}=e^{\eta}$, and used the obvious relation $d\eta=\xi^{-1}d\xi$
the microcanonical partition function of a system of interacting
particles may be rewritten as :
\begin{equation}
\Omega_{E,N}=\oint d\beta \oint d\xi \int\limits_{-\infty }^\infty
D\varphi \exp \left\{\beta E-N\ln \xi -\frac 1{2\beta }
\sum\limits_{s,s^{\prime }}\left( W_{ss^{\prime }}^{-1}\varphi
_s\varphi _{s^{\prime }}\right) \right\}\prod_{s}\left\{\xi
\exp(-\beta \varepsilon_{s}+\varphi_{s}) \right\}^{n_{s}}
\end{equation}
This microcanonical partition function can be used for
calculating all thermodynamically properties of the system with
fixed total number of particles and energy of system. Only in
this presentation can make summation over occupation number.
After summation over the occupation numbers $n_s$ \cite{Lev},
microcanonical partition function finally reduces to:
\begin{equation}
\Omega_{E,N}=\oint d\beta\oint d\xi \int\limits_{-\infty }^\infty
D\varphi \exp (\beta S_{eff}\left( \varphi ,\xi \right))
\end{equation}
where can introduce the effective entropy
\begin{equation}
S_{eff}\left( \varphi,\xi \right) =\frac{1}{2\beta
}\sum\limits_{s,s^{ \prime }}\left( W_{ss^{\prime }}^{-1}\varphi
_{s}\varphi _{s^{\prime }}\right) -\delta \sum\limits_{s}\ln
\left( 1+ \xi e^{\varphi-\beta \varepsilon _{s}}\right) +\left(
N+1\right) \ln \xi-\beta E
\end{equation}
where $\xi \equiv e^{\beta \mu}$ is absolute chemical activity of
chemical potential $\mu$. For this calculation was use the Fermi
statistic,because the two classical particle can not occupied one
spatial place. After this we can present partition function in
the usual form $\Omega_{E,N}=\exp S$, where $S$ is entropy of
system. For this goal, for determination microcanonical partition
function allows use the of efficient methods developed in the
quantum field theory without imposing additional restrictions of
integration over field variables or the perturbation theory. The
functional $\beta S_{eff}\left( \varphi,\lambda \right) $ depends
on distribution of the field variables $\varphi $ and the absolute
chemical activity $\xi$. The field variable $\varphi $ contains
the same information as original partition function with
summation of over occupation numbers, i.e. all information about
possible states of the systems. The saddle point method can now
be further employed to find the asymptotic value of the partition
function $\Omega_{E,N}$ for $N\rightarrow \infty $; the dominant
contribution is given by the states which satisfy the extreme
condition for the functional. The particles distribution is
determined by the saddle point solutions of equations:
\begin{equation}
\frac{\delta S_{eff}}{\delta \xi}=\frac{\delta S_{eff}}{\delta
\varphi }=0
\end{equation}
whether this distribution of particles is spatially inhomogeneous
or not. The solutions which correspond the finite entropy
$S_{eff}\left( \varphi,\lambda \right) $ while the volume of the
system tends to infinity, mean such solutions could be
thermodynamically stable. The above set of equations in principle
solves the many-particle problem in thermodynamic limit. The
spatially inhomogeneous solution of this equations corespondent
the distribution of interacting particles. Such inhomogeneous
behavior is associated with the nature and intensity of
interaction. In other words, accumulation of particles in a
finite spatial region (formation of a cluster) reflects the
spatial distribution of the fields and the activity. The inverse
matrix $\omega _{ss^{\prime }}^{-1}$ of the interaction, $\omega
_{ss^{\prime }}=\omega \left( \left|r_{s}-r_{s^{\prime }}\right|
\right) $, in continuum limit should be treated in the operator
sense \cite{Mag}, i.e.
\begin{equation}
\omega _{rr^{\prime }}^{-1}=\delta _{rr^{\prime
}}\widehat{L_{r^{\prime }}}=-\frac{1}{4\pi G m^{2}}\triangle
\end{equation}
where $m$ is the mass of particle and $\triangle $- Laplace
operator. With precision to surface term in continuum case the
effective entropy takes form:
\begin{equation}
S_{eff}\left( \varphi,\xi \right) =\int dV \left\{\frac{1}{8\pi G
m^{2}\beta}(\nabla \varphi)^{2}+\delta \sum_{p}\ln \left( 1-\delta
\xi e^{\varphi-\beta \varepsilon _{p}}\right)\right\} + N \ln
\xi-\beta E
\end{equation}
As shown before \cite{Bel}, in all cases classical Boltzmann
statistic for high temperature $\xi\leq 1$ and can use expansion
$\sum_{p}\ln \left( 1+\xi e^{\varphi-\beta \sum_{p}\varepsilon
_{p}}\right)\approx \xi e^{\varphi-\beta \varepsilon _{p}}+...$.
Integration over the impulse and coordinates should be performed
with regard for the cell volume $\left( 2\pi \hbar\right) ^{3}$
in the phase space of individual states \cite{Lev}. After
integration over the impulse the effective entropy can present in
general form
\begin{equation}
S_{eff}\left( \varphi,\xi \right) =\int dV \left\{\frac{1}{8\pi G
m^{2}\beta}(\nabla \varphi)^{2}-\xi A e^{\varphi}\right\}+N \ln
\xi-\beta E
\end{equation}
where $A \equiv \left(\frac{2\pi m}{\beta
\hbar^{2}}\right)^{\frac{3}{2}}$. The obtained microcanonical
partition function can use to calculation all thermodynamically
relation for self-gravitational system. At most general case it
is impossible, but for individual case can obtain the all
thermodynamic characteristic of system. Next will be examine
separate case.

First from all we consider the case of system noninteracting
particles, when $\varphi=0$. The effective entropy in this case
can write in the simple form
\begin{equation}
S_{eff}\left( \varphi,\xi \right) =-\int dV \xi \left(\frac{2\pi
m}{\beta \hbar^{2}}\right)^{\frac{3}{2}}+N\ln \xi-\beta E
\end{equation}
The extreme condition $\frac{\delta S_{eff}}{\delta
\xi}=\frac{\delta S_{eff}}{\delta \beta}=0$ reduce to two
equations
\begin{equation}
V \xi \left(\frac{2\pi m}{\beta \hbar^{2}}\right)^{\frac{3}{2}}=N
\end{equation}
and
\begin{equation}
V \xi \left(\frac{2\pi m}{\beta
\hbar^{2}}\right)^{\frac{3}{2}}=\frac{2}{3}\beta E
\end{equation}
from which at once determination the chemical activity
\begin{equation}
\xi =\frac{N}{V})\left(\frac{2\pi m}{\beta
\hbar^{2}}\right)^{-\frac{3}{2}}
\end{equation}
and well-know relation between fixed energy and number of
particles in system and inverse temperature
\begin{equation}
\beta=\frac{3}{2}\frac{N}{E}
\end{equation}
or $\frac{3}{2}NkT=E$. If determined coefficient substitute in to
effective entropy can obtain the ordinary entropy
\begin{equation}
S_{E,N}=\ln \frac{N!}{V}\left(\frac{4\pi mE}{3 N
\hbar^{2}}\right)^{\frac{3}{2}}+\frac{3N}{2}
\end{equation}
and the microcanonical partition function for fixed total energy
and number of not interacting particles, can write in the form
\begin{equation}
\Omega_{E,N}=\exp\left\{-\ln\frac{N!}{V}\left(\frac{3 N
\hbar^{2}}{4\pi mE}\right)^{\frac{3}{2}}+\frac{3N}{2}\right\}
\end{equation}
which reproduce well-know presentation microcanonical partition
function for ideal Boltzmann gas:
\begin{equation}
\Omega_{E,N}=\frac{V^N}{N!}\left\{\frac{4\pi m
Ee^{N}}{3N\hbar^{2}}\right\}^{\frac{3N}{2}}
\end{equation}
In the case homogeneous distribution particles $\nabla \varphi=0$
with average distance between them
$l=(\frac{V}{N})^{\frac{1}{3}}$ can introduce the average value
of potential
\begin{equation}
\varphi_{h}=\frac{3Gm^{2}N}{2E}(\frac{N}{V})^{\frac{1}{3}}
\end{equation}
and determine the ordinary entropy in the form
\begin{equation}
S_{E,N}=\ln \frac{N!}{Ve^{\varphi_{h}}}\left(\frac{4\pi mE}{3 N
\hbar^{2}}\right)^{\frac{3}{2}}+\frac{3N}{2}
\end{equation}
The microcanonical partition function can obtain in the form:
\begin{equation}
\Omega_{E,N}=\frac{V^N}{N!}e^{-N\widetilde{\varphi}}\left\{\frac{4\pi
m Ee^{N}}{3N\hbar^{2}}\right\}^{\frac{3N}{2}}
\end{equation}
Microcanonical partition function solve problem determination
thermodynamically properties of the self-gravitational system in
the case homogeneous distribution of particle. From entropy can
determine the pressure in self gravitational system from
homogeneous distribution of particle as $P_{h}=\frac{N}{\beta
Ve^{\varphi_{h}}}$ or equation of state in the form
\begin{equation}
P_{h}V_{h}=\frac{2E}{3e^{\varphi_{h}}}
\end{equation}

But in general case the distribution of particle in self-
gravitational system are inhomogeneous. Inhomogeneous
distribution of particle motivate the long-range gravitational
interaction. This system are unstable and all system divide as
finite state into few cluster finite size. The next task are in
developing possible method to determine partition function taking
into account the inhomogeneous distribution with formation
cluster of finite size. Let start with the situation when
homogeneous distribution particle decomposing to $n$ equal
cluster with average volume $V_{c}$. Inside this volume exist
nonhomogeneous distribution of particle with nonzero field
variable, and outside this volume the field variable are zero
because the particle are absence.

In this case the effective entropy can rewrite in the form:
\begin{equation}
S_{eff}\left( \varphi,\xi \right) =n \int^{V_{c}}_{0} dV
\left\{\frac{1}{4r_{m}}(\nabla \varphi)^{2}- \xi
Ae^{\varphi}\right\}-\xi A(V-nV_{c}) + N\ln \xi-\beta E
\end{equation}
where $r_{m}=2\pi Gm^{2}\beta $ and $A=\left( \frac{2\pi m}{\beta
h^{2}}\right)^{\frac{3}{2}}$ as previously. Minimization of
effective entropy on field variables lead to next saddle point
solutions of equation inside the cluster:
\begin{equation}
\frac{1}{2r_{m}}\Delta \varphi +\xi Ae^{\varphi}=0
\end{equation}
and normalization condition yields:
\begin{equation}
n \int^{V_{c}}_{0} dV \xi Ae^{\varphi}+\xi A (V-nV_{c})=N
\end{equation}
To multiply the first equation on $\nabla \varphi$ and used
relation $\Delta \varphi =\nabla (\nabla \varphi)$ can obtain the
first integral of this equation in the form:
\begin{equation}
\frac{1}{4r_{m}}(\nabla \varphi)^{2} +\xi Ae^{\varphi}=\Delta^{2}
\end{equation}
where $\Delta$ is unknown integral of "motion"  which must
determine from physical condition. If use the relation
$\beta\frac{d A}{d\beta}=-\frac{3}{2}A$ the saddle point equation
$\frac{\delta (S)}{\delta \beta}=0$ can rewrite in the form
\begin{equation}
n \int^{V_{c}}_{0} dV \left\{\frac{5}{2}\xi
Ae^{\varphi}-\Delta^{2}\right\}+\frac{3}{2}\xi A(V-nV_{c})=\beta E
\end{equation}
If introduce the density function in the form $\rho(r)\equiv \xi
A e^{\varphi}$ can rewrite the saddle point equations in the
simple form normalization condition
\begin{equation}
n\int dV \rho (r)+\xi A (V-nV_{c})=N
\end{equation}
and equation for conservation energy
\begin{equation}
n\int
dV\left\{\frac{5}{2}\rho(r)-\Delta^{2}\right\}+\frac{3}{2}\xi A
(V-nV_{c})=\beta E
\end{equation}
The usual entropy can present in the simple form
\begin{equation}
S=\frac{1}{2}N-2\beta E+N\ln\xi
\end{equation}
For presentation the entropy in term know parameter we must
determine unknown reverse temperature $\beta$ and chemical
activity $\xi$. The solution of obtained equation completely
solve problem statistical description of self-gravitational
system, but in general case this solutions are unknown. To make
an attempt to solve this problem in general case was take place
in article \cite{Lev}. Next will be present a certain solution of
self-gravitational system in the case formation few cluster same
size close pacing particle in ones. This condition correspond the
final state of self-gravitational system. The field variable
inside of cluster are constant and can be present as potential
between two close pacing particle $\varphi=\varphi_{0}=\frac{2\pi
Gm^{2}\beta}{2R}$ for $ r-r'= 2R$ where $R$ is size of particle.
For that is possible used the asymptotic value of field variable
in center of cluster and determine first integral as
$\Delta^{2}\equiv \xi A e^{\varphi_{0}}$. The initial size of
cluster can be determine from simple reason. In final case can
assume that all particle assemble only to $n$ clusters and take
into account the finite size of particle which occupied volume
$V_{0}=\frac{4\pi}{3}R^{3}$ can estimate $nV_{c}\simeq NV_{0}$.
The normalization condition in this case can present as
\begin{equation}
n\xi A e^{\varphi_{0}}+\xi A (V-nV_{c})=N
\end{equation}
and equation of conservation energy as
\begin{equation}
\frac{5}{2}n\xi A e^{\varphi_{0}}-nV_{c}\xi A
e^{\varphi_{0}}+\frac{3}{2}\xi A (V-nV_{c})=E
\end{equation}
from which can obtain the chemical activity
\begin{equation}
\xi A=\frac{N}{V-nV_{c}(1-e^{\varphi_{0}})}
\end{equation}
and relation $\beta E=\frac{3}{2}N$. The usual entropy can rewrite
in the other simple form
\begin{equation}
S=-N+N\ln\xi-\beta E=-N+N\ln\xi-\frac{3N}{2}
\end{equation}
Substitution obtained relation to effective entropy yields:
\begin{equation}
S = -N+ N \ln\frac{N}{A(V-nV_{c}(1-e^{\varphi_{0}}))}-\frac{3N}{2}
\end{equation}
This presentation fully solve problem the statistical description
of self gravitational system with formation few cluster equal
size. Using identity  $ N-Nln N\approx ln N!$ as result can
obtain the ordinary entropy in the form
\begin{equation}
S^{inh}_{E,N}=\ln \frac{V^N}{N!}\left\{\frac{4\pi m E e^{N}
}{3N\hbar^{2}(1-\frac{NV_{0}}{V}(1-e^{\varphi_{0}}))}\right\}^{\frac{3N}{2}}+\frac{3}{2}N
\end{equation}
and the partition function microcanonical ensemble for
self-gravitational system can present as:
\begin{equation}
Z_{E,N}=\frac{V^N}{N!}\left\{\frac{4\pi m E e^{N}
}{3N\hbar^{2}(1-\frac{NV_{0}}{V}(1-e^{\varphi_{0}}))}\right\}^{\frac{3N}{2}}
\end{equation}
If are not gravitational interaction between particles, than
$e^{\varphi_{0}}=1$ and the partition function reduces to
partition function of ideal Boltzmann gas of hard sphere. The
equation of state in the case inhomogeneous distribution of
particle (existence few cluster finite size) can present in the
form
\begin{equation}
P_{inh}V_{inh}=\frac{2E}{3(1-\frac{NV_{0}}{V}(1-e^{\varphi_{0}}))}
\end{equation}
After this calculation can contend that the entropy of
inhomogeneous distribution of particle (existence few cluster
finite size) is bigger as entropy homogeneous distribution of
particle
\begin{equation}
S_{inh}-S_{h}=\ln\frac{e^{\varphi_{h}}}{(1-\frac{NV_{0}}{V}(1-e^{\varphi_{0}}))}
\end{equation}
if $e^{\varphi_{h}}>1-\frac{NV_{0}}{V}(1-e^{\varphi_{0}})$ that
take place for real self gravitational system. The homogeneous
distribution of particle to meet the requirements of equilibrium
state.This is the thermodynamically reason formation inhomogeneous
distribution of particle. The relation between ordinary entropy
self gravitational system produce the next relation between
equation of state
\begin{equation}
\frac{P_{inh}V_{inh}}{P_{h}V_{h}}=\frac{e^{\varphi_{h}}}{(1-\frac{NV_{0}}{V}(1-e^{\varphi_{0}}))}>1
\end{equation}
If assume that the pressure in both states is equal, come to a
determination that the volume of inhomogeneous distribution of
particle of same self-gravitational system is bigger as volume of
homogeneous distribution of particle. If the domains is unlimited
the density of states diverges when the particles are spread to
infinity.. Therefore, there is no equilibrium state in strict
sense. Self-gravitational system have tendency to disperse. This
is already the case for ordinary gas in infinite volume. The
disperse rate is small in general and the system can be found in
quasi equilibrium state for a relatively long time.

Indeed,present equilibrium statistical description tell only
dilute structure in the self-gravitational system but not
describe meta stable states and tell nothing about time scales a
kinetic theory. The partition function have not any peculiarity
for different value of gravitational field. The problem of
description of the self-gravitational system of particles could
be solved with current approach where entropy for finite system
could be explicitly calculated. Spatial non-uniformity of
particles as the equilibrium state might alter necessary
activation barrier to proceed with transformation when the system
is being moved into non-equilibrium state. Gravity factor could
either promote or retard such transformation depending on the
system and conditions concerned.

\end{document}